\documentclass[prb,aps,twocolumn,showpacs,nofootinbib]{revtex4}

\usepackage{graphicx}
\usepackage{rotating}
\usepackage{amsmath}
\usepackage{amsfonts}
\usepackage{amssymb}
\usepackage[countmax]{subfloat}
\usepackage{dcolumn} % align table columns on decimal point
\usepackage{bm} % bold math
\usepackage{color}

\usepackage{float}
\usepackage{latexsym}

\begin{document}

\title{Dissipationless Spin Current between Two Coupled Ferromagnets}

%Magnetic Field Control of Dissipationless Spin Current

%Universal description of spin Josephson effect 

\author{Wei Chen$^{1}$, Peter Horsch$^{1}$, and Dirk Manske$^{1,2}$}

\affiliation{$^{1}$Max-Planck-Institut f$\ddot{u}$r Festk$\ddot{o}$rperforschung, Heisenbergstrasse 1, D-70569 Stuttgart, Germany
\\
$^{2}$Yukawa Institute for Theoretical Physics, Kyoto University, Kyoto 606-8502, Japan
}

\date{\rm\today}

\begin{abstract}

{We demonstrate the general principle which states that a dissipationless spin current flows between two coupled ferromagnets if their magnetic orders are misaligned. This principle applies regardless the two ferromagnets are metallic or insulating, and also generally applies to bulk magnetic insulators. On a phenomenological level, this principle is analogous to Josephson effect, and yields a dissipationless spin current that is independent from scattering. The microscopic mechanisms for the dissipationless spin current depend on the systems, which are elaborated in details. A uniform, static magnetic field is further proposed to be an efficient handle to create the misaligned configuration and stabilize the dissipationless spin current. }

\end{abstract}


% 85.75.-d spintronics

% 03.75.Lm Josephson effect quantum mechanics

% 75.10.Jm Quantized spin models

% 75.30.Ee Antiferromagnetics

\maketitle

\section{Introduction}

The basic principle of spintronics is to utilize spin current\cite{Maekawa13} to transport information. Various mechanisms such as spin field effect transistor (spin-FET)\cite{Datta90}, spin-transfer torque\cite{Berger96,Slonczewski96}, spin pumping\cite{Tserkovnyak02}, and (inverse) spin Hall effect\cite{Dyakonov71,Hirsch99,Saitoh06,Valenzuela06,Kimura07} have been proposed to generate, control, and detect the spin current, which made designing practical devices a reality. The spin current generated by these means is dissipative, i.e., a certain way of energy delivery is necessary to maintain the spin current. On the other hand, mechanisms for dissipationless spin transport have been proposed. For instance, a dissipationless spin current can be generated by an external electric field via intrinsic spin Hall effect\cite{Murakami03,Sinova04}. Another intensively studied example are the helical edge states in quantum spin Hall systems\cite{Kane05,Bernevig06,Bernevig06_2,Konig07} and topological superconductors\cite{Schnyder08,Kitaev09,Qi09}, which consist of dissipationless counterpropagate of opposite spins.

Another category of dissipationless spin transport is the spin Josephson effect proposed for various junctions\cite{Lee03,Nogueira04,Asano06,Linder07,Brydon08,Brydon09,Chasse10,Moor12,Brydon13} and measured in $^{3}$He-$B$ through narrow apertures\cite{BorovikRomanov88,BorovikRomanov89}. The general feature is that by coupling two systems that have $SO(3)$-breaking order parameters, a spontaneous spin current $J_{s}\propto\sin\theta$ flows across the junction, where $\theta$ is the angular difference of the two order parameters. A natural consequence of this spin supercurrent in junctions made of magnetic metals, in combination with the continuity equation, is the precession of the magnetization due to the fact that a spin supercurrent transfers angular momentum\cite{Slonczewski89,Brydon09,Chasse10}. This precession, which will be called "Josephson precession" throughout this article, serves as a new mechanism for the magnetic order parameters to spontaneously precess.

In this article, we demonstrate that a dissipationless spin current flows between two coupled ferromagnets when their magnetic order parameters are misaligned, regardless the two ferromagnets are metallic or insulating. We show that, as long as the coupling between the two ferromagnets has the generic form 
\begin{eqnarray}
H=J_{eff}{\bf S}_{L}\cdot{\bf S}_{R} 
\label{two_spins_Hamiltonian}
\end{eqnarray}
where ${\bf S}_{L/R}$ is the magnetic order parameter and $J_{eff}$ is the effective coupling, a dissipationless spin current occurs when ${\bf S}_{L}$ and ${\bf S}_{R}$ are misaligned. On a phenomenological level, the dissipationless spin current $J_{s}=-\partial E_{b}/\partial\theta$ can be derived from the commutation relation\cite{Villain74} $\left[\theta,S_{L}^{y}\right]=-\left[\theta,S_{R}^{y}\right]=i$, where $E_{b}=J_{eff}\langle S_{L}^{z}\rangle\langle S_{R}^{z}\rangle\cos\theta$ is the bound state energy stored in the system. This current-bound state relation is completely analogous to that in the Josephson effect, from which it is also concluded that this spin current is independent from scattering. On the microscopic level, the precise mechanism for the dissipationless spin current depends on the metallic or insulating nature of the two systems and their interface, which also determines if the dissipationless spin current is a equilibrium persistent current or a supercurrent carried by a certain condensate\cite{Sonin10}. These microscopic mechanisms will be elaborated in details. Besides two coupled bulk systems, this misalignment principle also applies to a bulk magnetic insulator, and accounts for the spin supercurrent described by the vector chirality ${\bf S}_{i}\times{\bf S}_{i+1}$\cite{Hikihara08,Okunishi08,Kolezhuk09}.

The central issue in experimentally realizing the dissipationless spin current is, of course, how to create the misaligned configuration between ${\bf S}_{L}$ and ${\bf S}_{R}$, especially the misalignment means that they are not in the energetically most favorable configuration (which would be parallel or antiparallel, depends on the sign of $J_{eff}$). This motivates us to study the effect of magnetic (${\bf B}$) field. We show that a uniform, static ${\bf B}$ field is an efficient handle to create the misalignment between magnetic order parameters. In systems where the effective coupling is antiferromagnetic (AF), $J_{eff}>0$, Larmor precession and Josephson precession are in opposite directions, hence the misalignment and the dissipationless spin current can be created by a ${\bf B}$ field in equilibrium, such as in a canted AF insulator. In addition, if the magnetization of a magnet is strongly pinned by shape anisotropy, one can coupled it to a misaligned magnetization by attaching it to a normal metal, and then use ${\bf B}$ field to magnetize the normal metal. An additional advantage of ${\bf B}$ field is that, a strong enough ${\bf B}$ field can fix the direction of magnetization against Josephson precession, hence stabilize the polarization of the dissipationless spin current. Moreover, applying ${\bf B}$ field in magnetic insulators with noncollinear order changes the vector chirality\cite{Hikihara08,Okunishi08,Kolezhuk09}, and yields a spin supercurrent that can flow out of the system. We give concrete examples for above mechanisms and estimate the realistic values of dissipationless spin current, together with experimental proposals to indirectly measure the dissipationless spin current.

The structure of the paper is the following. Sec.~II addresses the dissipationless spin current in the ferromagnetic metal/insulator/ferromagnetic metal (FMM/I/FMM) junction, and a corresponding two-spin phenomenological description. Sec.~III gives the mechanism for dissipationless spin current in the ferromagnetic metal/ferromagnetic insulator (FMM/FMI) junction. Sec.~IV addresses the FMI/superconductor/FMI (FMI/SC/FMI) junction and the related FMI/normal metal/FMI (FMI/N/FMI) junction. Sec.~V connects our analysis to the canted AF state due to magnetic field. The same analysis is further generalized to ferromagnetic (FM) or AF spirals in Sec.~VI. The results are summarized in Sec.~VII.

\section{FMM/I/FMM junction}

We now address FM metal/insulator/FM metal (FMM/I/FMM) junction as the first realization of the misalignment principle. This type of junction has been studied in Ref. \onlinecite{Nogueira04}, and reanalyzed in Ref.~\onlinecite{Chen13} in a different reference frame. We follow Ref.~\onlinecite{Chen13} since the relation between precession of the spins and the commutation relation is more transparent in this formalism. To be consistent with the rest of the paper, we choose the coordinate such that the relevant commutation relation is $\left[\theta,S^{y}\right]=i$, instead of $\left[\theta,S^{z}\right]=i$ in Refs.~\onlinecite{Nogueira04} and \onlinecite{Chen13}. Starting from the mean field treatment of Hubbard interaction in bulk FMM
\begin{eqnarray}
Un_{\uparrow}({\bf r})n_{\downarrow}({\bf r})
\rightarrow-c_{\uparrow}^{\dag}({\bf r})c_{\downarrow}({\bf r})\Delta({\bf r})-c_{\downarrow}^{\dag}({\bf r})c_{\uparrow}({\bf r})\Delta({\bf r})^{\dag}
\label{mean_field}
\end{eqnarray}
where $\Delta^{\dag}=U\langle S^{+}\rangle=U\langle S^{z}+iS^{x}\rangle=U\langle c_{\uparrow}^{\dag}c_{\downarrow}\rangle$, and we choose the spin quantization axis to be along ${\hat{\bf y}}$. The magnetization lies in the $S^{z}S^{x}$-plane with an angle $\theta$ that also determines the phase of $\Delta=|\Delta|e^{i\theta}$. We can directly utilize the formalism in Ref.~\onlinecite{Nogueira04} to calculate the spin current by setting $m=(2U/3)\langle S^{z}\rangle=0$ therein. Consider magnetizations on the two sides have the same magnitude $|\Delta_{L}|=|\Delta_{R}|=|\Delta|$ but are misaligned by an angle $\theta=\theta_{L}-\theta_{R}$. The spin supercurrent due to coherent tunneling of $\langle c_{\uparrow}^{\dag}c_{\downarrow}\rangle$ is 
\begin{eqnarray}
J_{s}^{y}&=&\frac{1}{2}\langle\dot{N}_{L\uparrow}-\dot{N}_{L\downarrow}\rangle=\frac{|T|^{2}}{2}S(0,|\Delta|)\sin\theta
\nonumber \\
&=&J_{s}^{y0}\sin\theta= N_{L}\langle \dot{S}_{L}^{y}\rangle=-N_{R}\langle \dot{S}_{R}^{y}\rangle
\label{spin_current_FIF}
\end{eqnarray}
where $|T|^{2}$ represents the tunneling amplitude, and $N_{L}$ and $N_{R}$ are total number of sites on each side of the junction. The function $S(a,b)$ satisfies $S(a,0)=0$ and $S(0,b)\neq 0$. By integrating Eq.~(\ref{spin_current_FIF}), the free energy of the junction is precisely described by the two-spin model in Eq.~(\ref{two_spins_Hamiltonian}), with $F=\langle H\rangle=J_{eff}\langle{\bf S}_{L}\cdot{\bf S}_{R}\rangle$, $J_{eff}=|T|^{2}S(0,|\Delta|)/2|{\bf S}_{L}||{\bf S}_{R}|$, and $J_{s}^{y}=-\partial F/\partial\theta$. This is analogous to the relation between Josephson current and Andreev bound state energy\cite{Zagoskin98} $I=-\partial E_{b}/\partial\phi$, where $\phi$ is the phase difference between the two superconductors. From the analogy to Josephson effect, it is also clear that this dissipationless spin current is an equilibrium current independent from scattering.

\begin{figure}[ht]
\begin{center}
\includegraphics[clip=true,width=0.95\columnwidth]{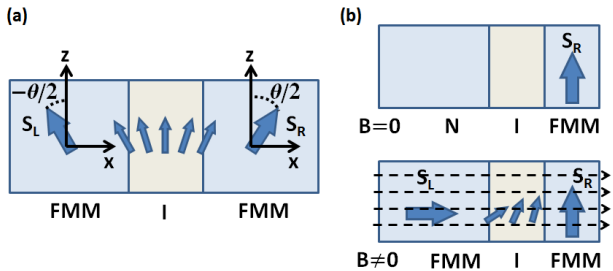}
\caption{(color online) (a) The FMM/I/FMM junction studied in Sec.~III. Blue arrows in the insulating region indicate proximity-induced magnetization. (b) Proposed realization of FMM/I/FMM junction. A normal metal is attached to a thin FMM film whose magnetization ${\bf S}_{R}$ is parallel to the interface because of strong shape anisotropy (top panel). In the presence of a ${\bf B}$ field that is smaller than the anisotropy field of the FMI, a magnetization ${\bf S}_{L}\parallel{\bf B}$ is induced in the normal metal (bottom panel). The misalignment can be adjusted by the ${\bf B}$ field.} 
\label{fig:FMMIFMM_junction}
\end{center}
\end{figure}

The physical picture of the spin supercurrent is more transparent within the superfluid description formulated in Ref.~\onlinecite{Chen13}. Within this description, the magnetization $\Delta\propto\psi=|\psi|e^{i\theta}$ is treated as a macroscopic condensate. The phase $\theta$ of the condensate is also the (local) angle of the magnetization, so the misaligned magnetizations in the two FMM's have different phases. Inside the insulator $0\leq x\leq L$, the proximity-induced magnetization gradually rotates to connect the two bulk magnetizations\cite{Tinkham96}
\begin{eqnarray}
\frac{\psi(x)}{\psi_{\infty}}=(1-\frac{x}{L})+\frac{x}{L}e^{i\theta}\;,
\label{gx_interface}
\end{eqnarray}
where $\psi_{\infty}$ is the bulk order parameter, as shown in Fig.~\ref{fig:FMMIFMM_junction}(a). This gradual rotation yields a gradient in $\psi(x)$ and hence the supercurrent in Eq.~(\ref{spin_current_FIF}). Thus the supercurrent can be viewed as originating from the proximity-induced phase gradient of the magnetization, and the supercurrent is carried by the magnetic condensate $|\psi|^{2}$, analogous to the charge supercurrent carried by Cooper pairs in a superconductor. The gradual rotation of magnetization should be measurable by polarization-sensitive probes such as optical Kerr effect\cite{Pechan05}, Lorentz transmission electron microscopy\cite{Graef00}, or magnetic transmission soft X-ray microscopy\cite{Fischer08}.

%We restrict our discussion to the case $|\Delta_{L}|=|\Delta_{R}|$, i.e., the two FMMs have the same magnitude of magnetization. If $|\Delta_{L}|\neq|\Delta_{R}|$, then this is similar to the conventional superconductor/insulator/superconductor(SIS) junction but for two superconductors that have different order parameters\cite{Chang94,Hurd96,Yip03,Wu04}, where more sophisticated treatment is necessary to clarify the current-phase relation, and we do not expect Eq.~(\ref{spin_current_FIF}) and the mapping to the two-spin model to be valid. However, our analysis is valid for if the two FMMs are made of the same material $|\Delta_{L}|=|\Delta_{R}|$ but of different size $N_{L}\neq N_{R}$. In such case, the two magnetizations still precess but around a tilted axis, and the Josephson spin current is still described by Eq.~(\ref{spin_current_FIF}). Because $N_{L}\langle S_{L}^{z}\rangle\neq N_{R}\langle S_{R}^{z}\rangle$ and they precess around a tilted axis, the net magnetization of the whole junction $N_{L}\langle {\bf S}_{L}\rangle+N_{R}\langle {\bf S}_{R}\rangle$ also precesses. The spin current in the lab observer's frame is an ac current similar to that described by Eq.~(\ref{ac_current}).

Since the magnitude of the order parameters $\langle S_{i}^{z}\rangle$ remains constant but $\langle S_{i}^{y}\rangle$ changes with time, the two order parameters are expected to precess around a common axis. The angle between $\langle{\bf S}_{L}\rangle$ and $\langle{\bf S}_{R}\rangle$ remains unchanged if there is no damping. This Josephson precession can be seen from the exact solution of the equation of motion derived from Eq.~(\ref{two_spins_Hamiltonian}): consider $\langle S_{L}^{z}\rangle=\langle S_{R}^{z}\rangle=S$ and defining the axis of precession as $z_{lab}$-axis, the solutions are $\langle S_{L}^{x}(t)\rangle_{lab}=S\sin(\theta/2)\cos\omega t$, $\langle S_{L}^{y}(t)\rangle_{lab}=S\sin(\theta/2)\sin\omega t$, and $\langle S_{L}^{z}(t)\rangle_{lab}=S\cos(\theta/2)$ from a lab observer's point of view, with angular velocity $\omega=2J_{eff}S\cos(\theta/2)$. In a realistic FMM/I/FMM junction, however, this Josephson precession is unlikely to be observed due to magnetic anisotropy or dipole-dipole interaction. To see this, one can estimate the spin supercurrent by assuming the junction size $\sim 1\mu$m$^{3}$,  $|T|\sim 1$meV, $\Delta\sim 10^{-1}$eV, density of states $N_{F}\sim 1/$eV. Adopting the formula for the critical charge current\cite{Ambegaokar63} but replace $2e\rightarrow\hbar$, one obtains $I_{s}^{0}=\hbar\left(\pi^{2}N_{F}^{2}\Delta|T|^{2}\overline{N}/\hbar\right)\sim  10^{15}\hbar/s$, where $\overline{N}\sim 10^{6}$ is number of cross section channels by assuming lattice constant $a\sim 1$nm. This yields $J_{eff}\sim 10^{-9}$eV for the corresponding two-spin model and the precession period $\sim 10^{-6}s$. However, the dipole-dipole energy for one spin on the right due to its neighboring spins is typically $U_{dip}^{intra}\sim 10^{-5}$eV$\gg J_{eff}$, which can easily suppress the precession. Moreover, it is unlikely to have a bulk magnetization that can remain isotropic down to energy scale $J_{eff}\sim 10^{-9}$eV, which is necessary for the precession. As a result, the order parameters remain static $\langle d{\bf S}_{i}/dt\rangle=0$. Nevertheless, the dissipationless spin current $J_{s}^{y}=-\partial F/\partial\theta$ due to the phase difference still occurs, although the torque it transfers is canceled by the dynamics due to anisotropy field, so the spins remain static.

A uniform ${\bf B}$ field can be used to induce the misalignment of magnetizations in a FMM/I/FMM junction. In the experiment proposed in Fig.~\ref{fig:FMMIFMM_junction}(b), a normal metal is attached to a thin FMM film whose magnetization ${\bf S}_{R}$ points parallel to the interface because of strong shape anisotropy. A ${\bf B}$ field smaller than the anisotropy field of the FMM is applied, which induces a magnetization ${\bf S}_{L}\parallel{\bf B}$ in the normal metal that is misaligned with ${\bf S}_{R}$. Both ${\bf S}_{L}$ and the misalignment angle $\theta$, and hence the magnitude of the spin supercurrent, can be arbitrarily adjusted by the ${\bf B}$ field. The spins remain static in equilibrium, but the dissipationless spin current $J_{s}^{y}=-\partial F/\partial\theta$ still occurs as long as ${\bf S}_{L}$ and ${\bf S}_{R}$ are misaligned. The torque transferred by the spin current is canceled out by the Larmor precession. 

%{\cblue Cite Loss-Balatsky paper here? Just say in 3D there's hint that it can also happen, although it's beyond our scope.}

%{\cblue Maybe cite Konig-MacDonald? Say it seems like this can happen inside a bulk too. But it's not clear to me if the superfluid description can work.}

\section{FMM/FMI junction}

In this section, we demonstrate that a FMM/FMI junction also has spontaneous dissipationless spin current when the magnetic moments are misaligned. It has been shown that in such a junction, magnons excited in the FMI transfers into the FMM as a dissipative spin current\cite{Takahashi09,Kajiwara10}. Here we consider a different situation that no magnons are excited in the FMI, but the its static magnetic moment is misaligned with that of the FMM. It will be shown that the Landau-Lifshitz type of dynamics at the interface causes a persistent spin current flowing into the FMM.

\begin{figure}[ht]
\begin{center}
\includegraphics[clip=true,width=0.95\columnwidth]{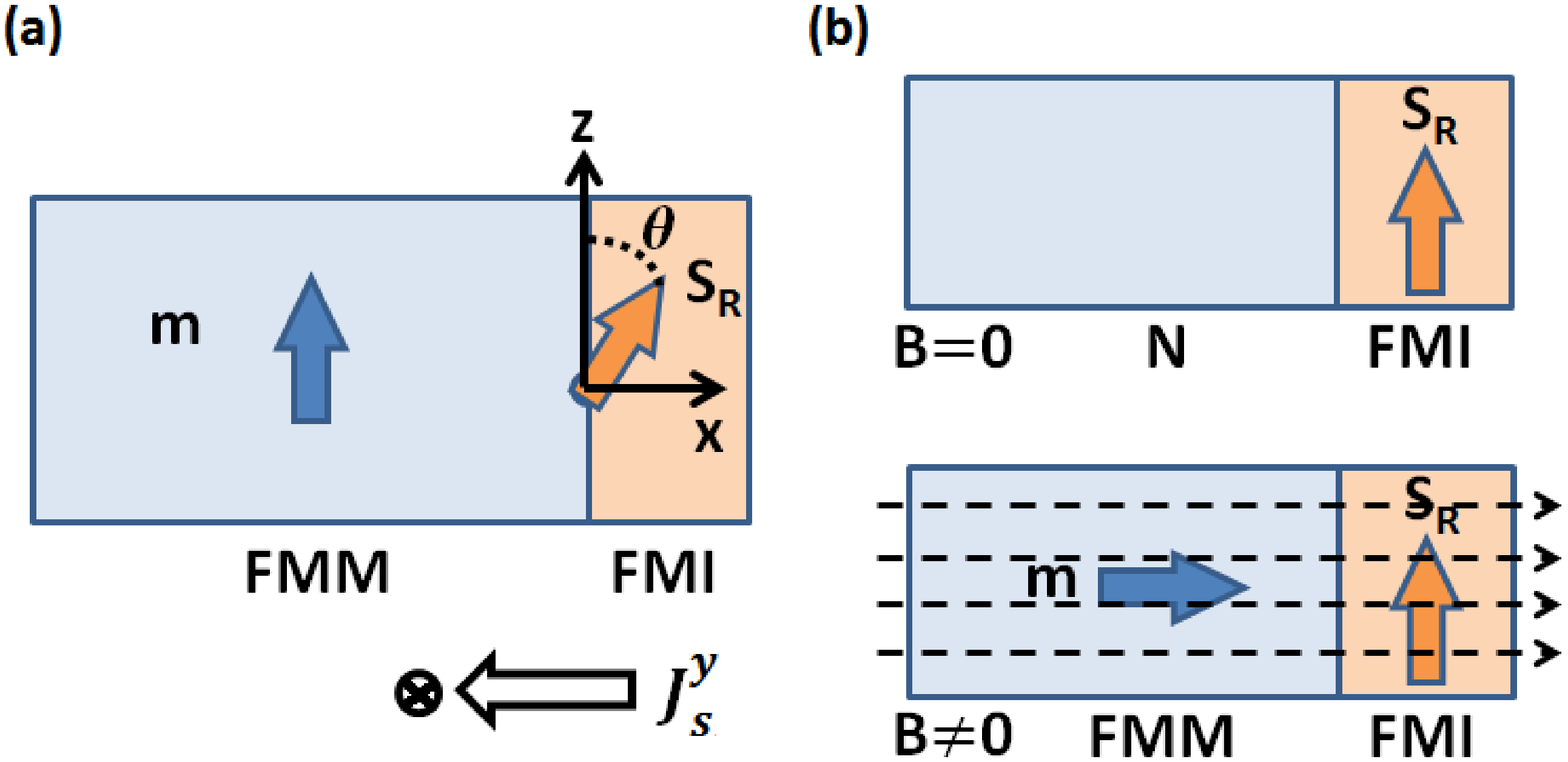}
\caption{(color online) (a) The FMM/FMI junction studied in Sec.~IV. The magnetization ${\bf m}$ of the FMM (blue arrow) is misaligned with magnetization ${\bf S}_{R}$ in the FMI (orange arrow) by an angle $\theta$. (b) Proposed realization of FMM/FMI junction. A normal metal is attached to a thin FMI film whose magnetization ${\bf S}_{R}$ is parallel to the interface because of strong shape anisotropy (top panel). In the presence of a ${\bf B}$ field that is smaller than the anisotropy field of the FMI, a magnetization ${\bf m}\parallel{\bf B}$ is induced in the normal metal (bottom panel). The misalignment can be adjusted by the ${\bf B}$ field. } 
\label{fig:FMM_FMI_junction}
\end{center}
\end{figure}

Consider a FMM occupying the space $-d_{S}\leq x\leq 0$, which interacts with an atomic layer of FMI at $x=0$ via a $s-d$ type interaction. Alternatively, the FMI can also occupy the whole $x>0$ space, which does not alter our conclusion. Denoting the momentum perpendicular and parallel to the interface to be $k_{x}$ and ${\bf k}_{\parallel}$, the Hamiltonian is 
\begin{eqnarray}
H&=&H_{0}+H_{I}\;,
\nonumber \\
H_{0}&=&\sum_{k_{x}{\bf k}_{\parallel}\sigma}\xi_{k_{x}{\bf k}_{\parallel}\sigma}c_{k_{x}{\bf k}_{\parallel}\sigma}^{\dag}c_{k_{x}{\bf k}_{\parallel}\sigma}
\nonumber \\
H_{I}&=&\Gamma{\bf S}_{R}\cdot \sum_{yz}c_{0yz\alpha}^{\dag}{\boldsymbol \sigma}_{\alpha\beta}c_{0yz\beta}
\nonumber \\
&=&\Gamma\left(\frac{a_{S}}{d_{S}}\right){\bf S}_{R}\cdot\sum_{k_{1x}k_{2x}{\bf k}_{\parallel}}{\bf m}_{k_{1x}k_{2x}{\bf k}_{\parallel}}
\end{eqnarray}
where $a_{S}$ is the lattice constant, $c_{0yz\alpha}$ is the electron annihilation operator at the interface ${\bf r}_{0}=(0,y,z)$, ${\bf m}_{k_{1x}k_{2x}{\bf k}_{\parallel}}=c_{k_{1x}{\bf k}_{\parallel}\alpha}^{\dag}{\boldsymbol\sigma}_{\alpha\beta}c_{k_{2x}{\bf k}_{\parallel}\beta}$, , $\xi_{k_{x}{\bf k}_{\parallel}\sigma}$ is the energy of electron with momentum $(k_{x},{\bf k}_{\parallel})$ and spin $\sigma$, and $\Gamma$ is the $s-d$ coupling. The spin current operator is 
\begin{eqnarray}
{\bf J}_{s}&=&\langle\hat{\bf J}_{s}\rangle=\frac{1}{2}\frac{d}{dt}\sum_{\bf k}\langle {\bf m}_{k_{x}k_{x}{\bf k}_{\parallel}}\rangle=\frac{i}{2}\langle\left[H_{I},\sum_{\bf k}{\bf m}_{k_{x}k_{x}{\bf k}_{\parallel}}\right]\rangle
\nonumber \\
&=&\Gamma\left(\frac{a_{S}}{d_{S}}\right)\sum_{k^{\prime}_{x}k_{x}{\bf k}_{\parallel}}\langle{\bf S}_{R}\times{\bf m}_{k^{\prime}_{x}k_{x}{\bf k}_{\parallel}}\rangle
\label{spin_current_operator}
\end{eqnarray}
Now consider the magnetization of FMM to be along $\hat{\bf z}$ direction ${\bf m}=\sum_{k_{x}{\bf k}_{\parallel}}{\bf m}_{k_{x}k_{x}{\bf k}_{\parallel}}=(0,0,m^{z})$, but the magnetization of the FMI points along a different direction ${\bf S}_{R}=\left(S_{R}^{x},S_{R}^{y},S_{R}^{z}\right)=\left(S\sin\theta,0,S\cos\theta\right)$, as shown in Fig.~\ref{fig:FMM_FMI_junction}(a). Equation (\ref{spin_current_operator}) then resembles a Landau-Lifshitz type of dynamics that is a natural consequence of the $s-d$ coupling and the misaligned magnetizations. In the configuration shown in Fig.~\ref{fig:FMM_FMI_junction}(a), only $J_{s}^{y}$ is nonzero
\begin{eqnarray}
J_{s}^{y}=-\Gamma\left(\frac{a_{S}}{d_{S}}\right)Sm^{z}\sin\theta
\label{Jy_FMMFMI}
\end{eqnarray}
which can be mapped into that produced by the two-spin model discussed in Sec.~II, 
\begin{eqnarray}
H_{eff}&=&J_{eff}\left({\bf S}_{R}\cdot{\bf S}_{m}\right)
\nonumber \\
\dot{{\bf S}}_{m}&=&J_{eff}\left({\bf S}_{R}\times{\bf S}_{m}\right)
\end{eqnarray}
where $J_{eff}=2\Gamma\left(a_{S}/d_{S}\right)$, ${\bf S}_{m}={\bf m}/2$, and $\dot{\bf S}_{m}$ is again calculated from the commutation relation $\left[\theta,S^{y}\right]=i$.

The effective coupling $J_{eff}$ apparently depends on the thickness $d_{S}$ of the FMM. For a FMM of size $d_{S}\sim\mu$m, $a_{S}\sim$nm, and $\Gamma\sim 0.1$eV, one obtains $J_{eff}\sim 10^{-4}$eV. Similar to the proposal in Sec.~III, a ${\bf B}$ field can be used to induce the misalignment of magnetizations in the FMM/FMI junction. As indicated in Fig.~\ref{fig:FMM_FMI_junction}(b), a normal metal is attached to a thin FMI film whose magnetization ${\bf S}_{R}$ is again pinned by strong shape anisotropy. A ${\bf B}$ field induces a magnetization ${\bf m}\parallel{\bf B}$ in the normal metal that is misaligned with ${\bf S}_{R}$.

%{\cred Besides several known mechanisms to produce spin supercurrent\cite{Sonin10}, our calculation shows that continued spin-flip of conduction electrons when they bounce from the interface can also produce a spin supercurrent that is a genuine persistent current. }

%Using linear response theory 
%\begin{eqnarray}
%J_{s}^{\alpha}=-i\int_{-\infty}^{t}dt^{\prime}\langle\left[\hat{J}_{s}^{\alpha}(t),H_{I}(t^{\prime})\right]\rangle
%\label{linear_response}
%\end{eqnarray}
%one can calculate the spin current flowing into the FMM.

\section{FMI/SC/FMI and FMI/N/FMI junction}

\begin{figure}[ht]
\begin{center}
\includegraphics[clip=true,width=0.55\columnwidth]{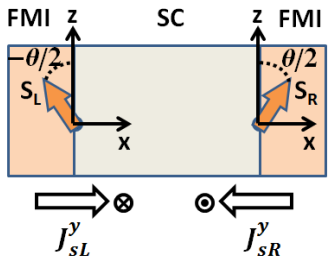}
\caption{ (color online) The FMI/SC/FMI junction studied in Sec. V. The dissipationless spin current $J_{sL}^{y}$ and $J_{sR}^{y}$ flowing into the SC from the two interfaces have opposite polarization.   } 
\label{fig:FMISCFMI_junction}
\end{center}
\end{figure}

Another realization of the two-spin model in Sec.~II is by sandwiching a type I s-wave SC between two FMI's (FMI/SC/FMI), as shown in Fig.~\ref{fig:FMISCFMI_junction}. The effect on SC has been considered long ago by de Gennes\cite{deGennes66}, and confirmed in the measurements of critical temperature reduction in FeNi/In/Ni\cite{Deutscher69} and Fe$_{3}$O$_{4}$/In/Fe$_{3}$O$_{4}$\cite{Hauser69} junctions. Following Ref. \onlinecite{deGennes66}, we consider the thickness of SC to be smaller than its coherence length $d_{S}<\xi_{0}$. The electrons in the last atomic layer of SC  interact with the two atomic layers of FMI's via $s-d$ coupling $\Gamma {\bf S}_{i}\cdot{\bf S}_{e}$, where ${\bf S}_{i}$ is the magnetization of the atomic layer per unit cell, $i=\left\{L,R\right\}$, and ${\bf S}_{e}$ is the electron spin. For $|{\bf S}_{L}|=|{\bf S}_{R}|=S$, the average ${\bf B}$ field as seen by the conduction electron in the SC region is then 
\begin{eqnarray}
\overline{h}(\theta)=2\Gamma S\left(\frac{a_{S}}{d_{S}}\right)\cos\frac{\theta}{2}\;,
\label{junction_average_field}
\end{eqnarray} 
where $a_{S}$ is the lattice constant of SC along the direction of the junction, and $\theta$ is the relative angle on $S^{x}S^{z}$-plane between magnetization on the left and right. The factor $a_{S}/d_{S}$ characterizes the extension of magnetic coupling at the FM/SC interface into the SC.  The condensation energy then changes according to\cite{deGennes66}
\begin{eqnarray}
E_{S}(\theta)-E_{N}&=&N(0)\left[-\frac{1}{2}\Delta^{2}+\overline{h}(\theta)^{2}\right]
\nonumber \\
&=&J_{eff}\langle{\bf S}_{L}\cdot{\bf S}_{R}\rangle+const.
\label{condensation_energy}
\end{eqnarray}
where $N(0)$ is the density of states at the Fermi surface, $\Delta$ is the SC gap, and $J_{eff}=2|\Gamma|^{2}\left(a_{S}/d_{S}\right)^{2}N(0)$ is the effective AF coupling after mapping to the two-spin model described by Eq.~(\ref{two_spins_Hamiltonian}). Eq. (\ref{condensation_energy}) is valid for $h_{0}=\overline{h}(\theta=0)<\Delta/\sqrt{2}$. For junctions with $h_{0}>\Delta/\sqrt{2}$, there exists a critical angle $\theta_{c}$ beyond which SC is destroyed,\cite{deGennes66} so Eq. (\ref{condensation_energy}) is valid only for $\theta<\theta_{c}$ in this case.

%In reality, it is perhaps easier to observe this precession if one side of the junction is much thicker $L\gg a$ such that its first atomic layer ${\bf S}_{L}$ is strongly coupled to the bulk and does not precess. This side can serve as the substrate too. The other side ${\bf S}_{R}$ then precesses around ${\bf S}_{L}$ with a speed twice of Eq. (\ref{junction_spin_current}). Such precession can be measured by usual technique such as ferromagnetic resonance(FMR). We expect the bound state energy will eventually be dissipated, which can be modeled by adding a Gilbert damping term in Eq. (\ref{two_spins_EOM}).  However, Eq. (\ref{junction_average_field}) implies an effective AF coupling $J_{eff}>0$, so one can apply a uniform magnetic field to stop the precession and produce a damping-free, field-adjustable persistent dc spin current according to Eq. (\ref{eq_angle}). This serves as a novel mechanism to maintain a dc spin current between two magnets. 

%Eq. (\ref{condensation_energy}) also applies if the SC is replaced by a normal metal (FMI/N/FMI junction), provided that the wave function of the electrons in the normal metal remains coherent. In such case, the bound state energy is just the magnetic potential energy due to polarization of the normal metal, i.e. $E_{N}=-N(0)\overline{h}(\theta)^{2}$ in Eq. (\ref{condensation_energy}), which is negative. This indicates an effective FM coupling $J_{eff}=-2|\Gamma|^{2}\left(a_{S}/d_{S}\right)^{2}N(0)<0$ between the two FMI's. 

We shall prove now that a dissipationless spin current flows through the SC when the magnetizations in the two FMI's are misaligned. Denoting the momentum parallel to the interface to be ${\bf k}_{\parallel}$ and assuming the energy along the junction is quantized and labeled by $n$, the argument in Ref.~\onlinecite{deGennes66} implies the coupling at the two interfaces extends into the bulk SC
\begin{eqnarray}
H_{I}&=&\Gamma\sum_{yz}\left[{\bf S}_{L}\cdot c_{0yz\alpha}^{\dag}{\boldsymbol\sigma}_{\alpha\beta}c_{0yz\alpha}+
{\bf S}_{R}\cdot c_{Nyz\alpha}^{\dag}{\boldsymbol\sigma}_{\alpha\beta}c_{Nyz\alpha}\right]
\nonumber \\
&=&\Gamma\left(\frac{a_{S}}{d_{S}}\right)\left({\bf S}_{L}+{\bf S}_{R}\right)\cdot\sum_{n{\bf k}_{\parallel}}{\bf m}_{nn{\bf k}_{\parallel}}
\label{H_FMISCFMI}
\end{eqnarray}
where $c_{0yz\alpha}$ and $c_{Nyz\alpha}$ are electron annihilation operators at the left and right interfaces, respectively. $c_{n{\bf k}_{\parallel}\alpha}$ is the electron annihilation operator of energy level $n$ and momentum ${\bf k}_{\parallel}$, and ${\bf m}_{nn{\bf k}_{\parallel}}=c_{n{\bf k}_{\parallel}\alpha}^{\dag}{\boldsymbol\sigma}_{\alpha\beta}c_{n{\bf k}_{\parallel}\beta}$. The spin current operator is, following Eq.~(\ref{spin_current_operator}), 
\begin{eqnarray}
\hat{J}_{s}^{\alpha}&=&\frac{1}{2}\frac{d}{dt}\sum_{n{\bf k}_{\parallel}}m_{nn{\bf k}_{\parallel}}^{\alpha}=\hat{J}_{sL}^{\alpha}+\hat{J}_{sR}^{\alpha}
\nonumber \\
\hat{J}_{si}^{\alpha}&=&\Gamma\left(\frac{a_{S}}{d_{S}}\right)\sum_{n{\bf k}_{\parallel}}\left[{\bf S}_{i}\times{\bf m}_{nn{\bf k}_{\parallel}}\right]^{\alpha}
\end{eqnarray}
where $\hat{J}_{sL}^{\alpha}$ and $\hat{J}_{sR}^{\alpha}$ are contributions to the spin current due to coupling to the left and right FMI, respectively. Consider $|{\bf S}_{L}|=|{\bf S}_{R}|$, then from the symmetry of the problem one immediately concludes that $\hat{J}_{sL}^{\alpha}=-\hat{J}_{sR}^{\alpha}$, so the net current flowing into the SC is zero, as indicated in Fig.~\ref{fig:FMISCFMI_junction}. Nevertheless, in what follows we show that each of $\hat{J}_{sL}^{\alpha}$ and $\hat{J}_{sR}^{\alpha}$ is nonzero, and their opposite signs mean that current flowing in from the left interface immediately flows out through the right interface, so continuity equation is satisfied.

The spin current $J_{sL}^{y}$ contributed from ${\bf S}_{L}$ alone can be calculated by linear response theory 
\begin{eqnarray}
J_{sL}^{y}&&=-i\int_{-\infty}^{t}dt^{\prime}\langle\left[\hat{J}_{sL}^{y}(t),H_{I}(t^{\prime})\right]\rangle
\nonumber \\
&&=i\Gamma^{2}\left(\frac{a_{S}}{d_{S}}\right)^{2}
S_{L}^{x}\left(S_{L}^{z}+S_{R}^{z}\right)
\nonumber \\
&&\times\int_{-\infty}^{t}dt^{\prime}\sum_{n{\bf k}_{\parallel}}\sum_{l{\bf k}^{\prime}_{\parallel}}\langle\left[m_{nn{\bf k}_{\parallel}}^{z}(t),m_{ll{\bf k}^{\prime}_{\parallel}}^{z}(t^{\prime})\right]\rangle
\label{JsL_linear_response}
\end{eqnarray}
The integration is the susceptibility $\chi^{zz}_{0}=-2N(0)$ followed by Eq.~(\ref{condensation_energy}), thus
\begin{eqnarray}
J_{sL}^{y}&=&-2\Gamma^{2}\left(\frac{a_{S}}{d_{S}}\right)^{2}
S_{L}^{x}\left(S_{L}^{z}+S_{R}^{z}\right)N(0)
\nonumber \\
&=&2\Gamma^{2}\left(\frac{a_{S}}{d_{S}}\right)^{2}
S^{2}\sin\theta N(0)
\label{spin_current_FMINFMI}
\end{eqnarray}
Note that $J_{sL}^{y}>0$ (and hence $J_{sR}^{y}<0$) is consistent with the claim that the two FMI's are coupled antiferromagnetically, and the effective coupling $J_{eff}=2|\Gamma|^{2}\left(a_{S}/d_{S}\right)^{2}N(0)$ is recovered.

One can view the FMI/SC/FMI junction as a combination of two FMM/FMI junctions mentioned in Sec.~IV. This is because under the proximity-induced ${\bf B}$ field, Eq.~(\ref{junction_average_field}), the SC is magnetized by $m^{z}=\chi_{0}^{zz}\overline{h}(\theta)=-4\Gamma S\left(a_{S}/d_{S}\right)N(0)\cos\left(\theta/2\right)$. The $s-d$ coupling between $m^{z}$ and one of the FMI's, say ${\bf S}_{L}$, yields a Landau-Lifshitz type of dynamics described by Eqs.~(\ref{spin_current_operator}) and (\ref{Jy_FMMFMI}), from which the $J_{sL}^{y}$ in Eq.~(\ref{spin_current_FMINFMI}) is recovered. Thus the dissipationless spin current is a combined effect of proximity-induced magnetization and the Landau-Lifshitz dynamics due to the $s-d$ coupling.

The analysis in Eqs.~(\ref{H_FMISCFMI})$\sim$(\ref{spin_current_FMINFMI}) also applies if the SC is replaced by a normal metal (FMI/N/FMI junction). In such case, the effective coupling between the two FMI's is described by the Ruderman-Kittel-Kasuya-Yosida (RKKY) interaction calculated for two parallel walls of distance $r_{LR}$, $J_{eff}=-\Gamma^{2}F(r_{LR})$. Despite the oscillation of effective coupling with $r_{LR}$, one reaches the same generic conclusion: a spin current flows between two FMI's coupled via RKKY interaction, as long as their magnetizations are misaligned.

The two $\langle {\bf S}_{i}\rangle$'s are expected to precess if they are prepared to be non-collinear, again due to the commutation relation $\left[\theta,S^{y}\right]=i$. This precession is, however, unlikely to be observed in a realistic FMI/SC/FMI junction. Assuming exchange coupling to be $|\Gamma|\sim 0.1$eV, and $N(0)\sim 1/$eV, and a junction of size $d_{S}\sim \mu$m and lattice constant $a_{S}\sim n$m. They are roughly the parameters extracted from the experiment on Fe$_{3}$O$_{4}$/In/Fe$_{3}$O$_{4}$ junction\cite{Hauser69}. They correspond to a two-spin model with effective coupling $J_{eff}\sim 10^{-8}$eV, and one needs only an external field $|{\bf B}_{ext}|\sim 1$G to make a significant angle $\theta$ between magnetization in the two FMIs. The dipole-dipole interaction for one spin in the left bulk due to the spins on its neighboring sites also in the left bulk is $U_{dip}^{intra}\sim 10^{-5}$eV$\gg J_{eff}$, which easily destroys the Josephson precession. Moreover, in a bulk FMI, it is unlikely that the spins can remain isotropic down to energy scale $J_{eff}\sim 10^{-8}$eV. Therefore the Josephson precession is unlikely to be observed. We close this section by remarking that the two experiments on FMI/SC/FMI junction already demonstrated the advantage of magnetic field\cite{Hauser69,Deutscher69}. The $T_{c}$ reduction of the SC layer as a function of orientation of magnetization in the two FMIs, controlled by strong magnetic field $g\mu_{B}|{\bf B}|\gg J_{eff}$, justifies the condensation energy in Eq. (\ref{condensation_energy}) proposed by de Gennes, and also serves as an indirect evidence for the dissipationless spin current. 

%{\cred Our claim is then the FMI/SC/FMI or FMI/N/FMI junction contains a spin supercurrent described by Eq.~(\ref{spin_current_FMINFMI}), which may be measurable by inverse spin Hall effect\cite{Saitoh06,Valenzuela06,Kimura07} if the SC or normal metal has strong spin-orbit coupling, where the voltage developed in the transverse direction can be measured by a high impedance voltmeter. }

%Thus magnetic field is an efficient handle to control $T_{c}$ in a FMI/SC/FMI junction, which may be used as a device to control the supercurrent in a superconducting thin film. 

%{\cred The physical picture for this spin supercurrent is more clear if one defines the spin quantization axis to be along ${\hat{\bf y}}$. One observes that $m^{z}\propto\langle c_{\uparrow}^{\dag}c_{\uparrow}-c_{\downarrow}^{\dag}c_{\downarrow}\rangle$ defined for spin quantized along ${\hat{\bf z}}$ is equivalent to $m^{z}=\left(m^{+}+m^{-}\right)/2\propto\langle c_{\uparrow^{\prime}}^{\dag}c_{\downarrow^{\prime}}+c_{\downarrow^{\prime}}^{\dag}c_{\uparrow^{\prime}}\rangle$ defined for spins quantized along ${\hat{\bf y}}$, denoted by $\uparrow^{\prime}$ and $\downarrow^{\prime}$. The Green's functions are connected by spin-flip processes at the interface for spins quantized along ${\hat{\bf y}}$, yielding a correlation function that contributes to the spin supercurrent, as shown in Fig.~\ref{fig:FMINFMI_junction}(b). This continued spin-flip at the interface is the origin for the spin supercurrent with polarization ${\boldsymbol \mu}\parallel{\hat{\bf y}}$. } 

\section{Canted Antiferromagnetic State}

Next, we demonstrate that the AF state of localized spins on a bipartite lattice is also an example of the misalignment principle. The insight comes from the consideration of two antiferromagnetically coupled atomic spins. Since they are precisely described by Eq. (\ref{two_spins_Hamiltonian}), mapping to the two-spin model is trivial. If the exchange coupling stems from a two-site Hubbard model in the $U\gg t$ limit, it is shown that the corresponding Heisenberg model with $J=4t^{2}/U$ has a spin supercurrent ${\bf J}_{s}=J\langle{\bf S}_{L}\times{\bf S}_{R}\rangle$ flowing between the two sites due to virtual hopping of electrons, if the two spins are prepared to be misaligned\cite{Bruno05}. Or in a more realistic model that contains virtual hopping of electrons between two transition metal ions and an oxygen sandwiched in between, one reaches similar conclusions\cite{Katsura05}. These mechanisms that involve virtual hopping of electrons provide a microscopic origin for the interatomic spin supercurrent.

%What is not emphasized in Ref.~\onlinecite{Bruno05} is the precession of the two atomic spins due to angular momentum conservation, and polarization of the spin current constantly changes with time as formulated in Eq.~(\ref{ac_current}).  

Generalizing this two-site model to bulk AF state is straight forward. Consider the Heisenberg model defined on a $d$-dimensional square lattice
\begin{eqnarray}
H=\sum_{\langle ij\rangle}J{\bf S}_{i}\cdot{\bf S}_{j}
\end{eqnarray}
with $J>0$. Define the spins on $A$ sublattice to be ${\bf S}_{A}$ and $B$ sublattice to be ${\bf S}_{B}$. The spin supercurrent flowing between $A$ and $B$ when $\langle{\bf S}_{A}\rangle$ are prepared to be in a certain direction and $\langle{\bf S}_{B}\rangle$ in a different direction is 
\begin{eqnarray}
\langle \dot{S}_{A}^{y}\rangle=zJ\langle S_{A}^{z}\rangle
\langle S_{B}^{z}\rangle\sin\theta
=-\langle \dot{S}_{B}^{y}\rangle
\label{two_spins_time_evolution_cantedAF}
\end{eqnarray}
where $z$ is number of nearest neighbor sites that belong to a different sublattice ($z=4$ for 2D square lattice). Therefore all the $A$ sites have spin supercurrent flowing in from the $z$ neighboring $B$ sites, again due to virtual hopping of electrons on each bond, and all $B$ sites have spin supercurrent flowing out to the $z$ neighboring $A$ sites. Note that for a lattice constant $a\sim 1$nm, the dipole-dipole interaction between localized $U_{dip}^{intra}\sim 10^{-5}$eV is much smaller than the AF coupling $J\sim 0.1$eV, so it can be safely ignored. The spins should precess as long as no significant anisotropy is present.

\begin{figure}[ht]
\begin{center}
\includegraphics[clip=true,width=0.95\columnwidth]{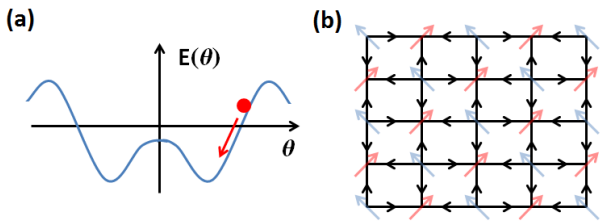}
\caption{ (color online) (a) Schematics of the classical energy, Eq. (\ref{Etotal_magnetic_field_SS}). Whatever is the initial configuration of the spins (red dot), the system relaxes to the energy minimum through damping, in which a pattern of spin supercurrent persists. (b) Pattern of spin current (black arrows) of a field-induced canted AF state on a 2D square lattice. Spins on the two sublattices, $\langle{\bf S}_{A}\rangle$ and $\langle{\bf S}_{B}\rangle$, are labeled by red and blue arrows, respectively. The polarization of the spin current is parallel to $\langle{\bf S}_{A}\times{\bf S}_{B}\rangle$, which is determined by the magnetic field. } 
\label{fig:AF_spin_current_pattern}
\end{center}
\end{figure}

Since the coupling is AF, applying a magnetic field changes the ground state configuration. The bound state energy per two sites in the presence of a magnetic field is
\begin{eqnarray}
E_{total}=zJ_{eff}S^{2}\cos\theta-2g\mu_{B}BS\cos\frac{\theta}{2}\;,
\label{Etotal_magnetic_field_SS}
\end{eqnarray}
where $S=\langle S_{A}^{z}\rangle=\langle S_{B}^{z}\rangle$. The angle that minimizes the energy is
\begin{eqnarray}
\theta_{0}=2\arccos\left(\frac{g\mu_{B}B}{2zJ_{eff}S}\right)\;,
\label{eq_angle}
\end{eqnarray}
between spins on $A$ sites and spins on $B$ sites. This is precisely the spin configuration calculated from the classical energy in the canted AF state\cite{Weihong91,Hamer92,Igarashi92,Yang97,Gluzman93,Zhitomirsky98}, if one ignores higher order contributions from quantum fluctuations. The purpose of our simple analysis is not to capture the precise value of the canting angle and magnetization, but to show that this canting angle yields a spin supercurrent flowing between $A$ and $B$ sites according to Eq.~(\ref{two_spins_time_evolution_cantedAF}). The system remains equilibrium and no precession occurs, because Larmor precession and Josephson precession are in opposite direction and canceled each other. Whatever is the initial angle, the system releases to $\theta_{0}$ through damping, and then the spins remain static in this energy minimum, as indicated in Fig. \ref{fig:AF_spin_current_pattern}(a). The pattern of spin supercurrent is shown in Fig.~\ref{fig:AF_spin_current_pattern}(b) for a canted AF state on a 2D square lattice. Despite such a pattern of spin supercurrent, which should extend up to correlation length, the alternating signs of the current on the lattice implies that it is averaged out and unlikely to be observable in the mesoscopic scale.

\section{coplanar magnets with noncollinear order }

%Previous investigations on frustrated spin chains have already shed the light on this issue\cite{Hikihara08,Okunishi08,Kolezhuk09}. In intrinsically frustrated models such as $J_{1}-J_{2}$ model, applying a magnetic field affects the vector chirality \begin{eqnarray}{\boldsymbol\kappa}_{ij}=\langle{\bf S}_{i}\times{\bf S}_{j}\rangle\label{vector_chirality}\end{eqnarray}that is directly proportional to the spin current between site $i$ and $j$. Assuming quantum coherence maintains up to mesoscopic scale, this spin current should be measurable by detecting the voltage drop due to the induced electric dipolar field\cite{Meier03}.

The discussion about the canted AF state motivates us to search for systems where the field-induced interatomic spin supercurrent can be measured on a mesoscopic scale. In this section, we argue that for coplanar magnets with noncollinear orders, the spin current induced by an in-plane magnetic field is a generic feature independent from many details such as dimensionality, magnitude of the spin, and sign of exchange coupling. These details only affect the magnitude and pattern of the spin current. Our insight comes from the gauge interpretation that connects noncollinear magnets due to Dzyaloshinskii-Moriya(DM) interaction to spin Josephson effect\cite{Katsura05}. The angle between neighboring spins in the noncollinear ground state is equivalent to an extra phase in the spin Josephson effect. Applying a magnetic field changes the angle between neighboring spins and hence creates a spin supercurrent that can flow out of the system.

%We remark that for chiral helimagnets of localized spins, applying a magnetic field perpendicular to the helical axis yields a kink crystal whose motion creates a spin current\cite{Bostrem08,Bostrem08_2}. This is not included in the much simpler mechanism we present here.   

\begin{figure}[ht]
\begin{center}
\includegraphics[clip=true,width=0.95\columnwidth]{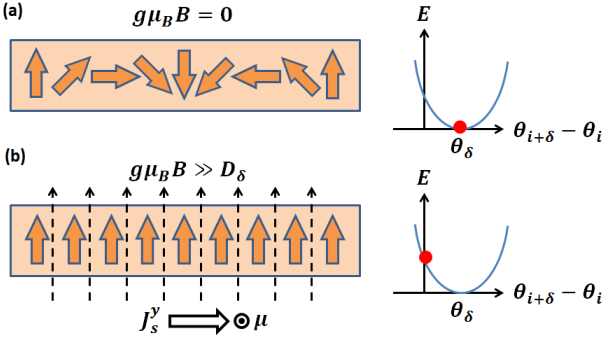}
\caption{(color online) (a) A coplanar spiral order at zero field (left panel) and schematics of the ground state (red point) in the classical energy versus neighboring angle plot (right panel). (b) The strong in-plane magnetic field $g\mu_{B}B\gg D_{\delta}$ forces the spins to point at the same direction (left panel). The lifting of ground state energy (right panel) creates a uniform spin current $J_{s}=D_{\delta}S^{2}$ with out-of-plane polarization ${\boldsymbol\mu}$. } 
\label{fig:spiral_at_large_B}
\end{center}
\end{figure}

We consider the coplanar spiral due to DM interaction in either 1D chain or 2D square lattice
\begin{eqnarray}
H=\sum_{i,\delta}J{\bf S}_{i}\cdot{\bf S}_{i+\delta}-{\bf D}_{\delta}\cdot\left({\bf S}_{i}\times{\bf S}_{i+\delta}\right)
\label{DM_Hamiltonian}
\end{eqnarray}
where ${\boldsymbol\delta}=\left\{{\bf a},{\bf b}\right\}$ for 2D square lattice with lattice constants ${\bf a}$ and ${\bf b}$, and ${\boldsymbol\delta}={\bf a}$ in 1D. First we consider FM spiral $J<0$, and we choose ${\bf D}_{\delta}=D_{\delta}{\bf {\hat y}}$ with $D_{\delta}>0$. Eq. (\ref{DM_Hamiltonian}) can be written as
\begin{eqnarray}
H=\sum_{i,\delta}\frac{\tilde{J}_{\delta}}{2}\left(e^{i\theta_{\delta}}S_{i}^{+}S_{i+\delta}^{-}
+e^{-i\theta_{\delta}}S_{i}^{-}S_{i+\delta}^{+}\right)+JS_{i}^{y}S_{i+\delta}^{y}\;,
\nonumber \\
\label{DM_Hamiltonian_2}
\end{eqnarray}
where $\tilde{J}_{\delta}=-\sqrt{J^{2}+D_{\delta}^{2}}$, $\sin \theta_{\delta}=D_{\delta}/|\tilde{J}_{\delta}|$, and the spin flip operators are defined as $S_{i}^{\pm}=S_{i}^{z}\pm iS_{i}^{x}$. By writing the spin configuration as $\langle{\bf S}_{i}\rangle=S\left(\sin\theta_{i},0,\cos\theta_{i}\right)$, the classical energy is $E=\sum_{i,\delta}{\tilde J}S^{2}\cos\left(\theta_{i+\delta}-\theta_{i}-\theta_{\delta}\right)$. The ground state configuration is a coplanar spiral where spins lie on $xz$-plane with neighboring angle $\theta_{i+\delta}-\theta_{i}=\theta_{\delta}={\bf Q}\cdot{\boldsymbol\delta}$, where ${\bf Q}$ is the spiral wave vector, as indicated in Fig.~\ref{fig:spiral_at_large_B}(a).

In the presence of an in-plane magnetic field ${\bf B}\perp{\bf D}$, the classical energy is modified by
\begin{eqnarray}
E=\sum_{i,\delta}{\tilde J}_{\delta}S^{2}\cos\left(\theta_{i+\delta}-\theta_{i}-\theta_{\delta}\right)-\sum_{i}g\mu_{B}BS\cos\theta_{i}\;.
\nonumber \\
\label{FM_spiral_energy_with_B_field}
\end{eqnarray}
The spin supercurrent due to virtual hopping of the electrons are known to be the microscopic mechanism for the DM interaction and cause the magnetic spiral\cite{Bruno05,Katsura05}. However, this spin supercurrent cannot flow out of the system and is not what we concern in this section. Instead, we consider the fact that magnetic field lifts the ground state energy of the spiral and yields a spin supercurrent {\it that can flow out of the system}
\begin{eqnarray}
J_{s,i,i+\delta}^{y}&=&-\frac{\partial E}{\partial (\theta_{i+\delta}-\theta_{i})}=-|\tilde{J}|S^{2}\sin\left(\theta_{i+\delta}-\theta_{i}-\theta_{\delta}\right)
\nonumber \\
&&\label{spin_current_FM_spiral}
\end{eqnarray}
since the spin dynamics is still governed by the commutation relation $\left[\theta,S^{y}\right]=i$. The extra phase $\theta_{\delta}$ means that the spin current that can flow out of the system is proportional to the vector chirality defined in a rotated reference frame
\begin{eqnarray}
J_{s,i,i+\delta}^{y}\propto |{\boldsymbol\kappa}^{\prime}_{i,i+\delta}|=|\langle{\bf S}_{i}^{\prime}\times{\bf S}_{i+\delta}^{\prime}\rangle|\;,
\label{chirality_rotated}
\end{eqnarray} 
where
\begin{eqnarray}
S_{i}^{\prime z}&=&S_{i}^{z}\cos{\bf Q}\cdot{\bf r}_{i}+S_{i}^{x}\sin{\bf Q}\cdot{\bf r}_{i}\;,
\nonumber \\
S_{i}^{\prime x}&=&-S_{i}^{z}\sin{\bf Q}\cdot{\bf r}_{i}+S_{i}^{x}\cos{\bf Q}\cdot{\bf r}_{i}\;,
\nonumber \\
S_{i}^{\prime y}&=&S_{i}^{y}\;.
\end{eqnarray}
The polarization of the spin current points out of plane ${\boldsymbol\mu}\parallel{\hat{\bf y}}$. Eq.~(\ref{chirality_rotated}) is similar to that defined for systems without DM interaction\cite{Hikihara08,Okunishi08,Kolezhuk09}, except the extra phase in Eq.~(\ref{spin_current_FM_spiral}) requires the vector chirality to be defined in a rotated reference frame. The spin current that can flow out of the sample is zero\cite{Katsura05} when $B=0$ because $\theta_{i+\delta}-\theta_{i}=\theta_{\delta}$, or equivalently ${\boldsymbol\kappa}_{i,i+\delta}^{\prime}=0$. A magnetic field creates a spin current simply because its presence modifies the angle of each spin, such that $\theta_{i+\delta}-\theta_{i}\neq\theta_{\delta}$ and consequently the rotated vector chirality $\langle{\bf S}_{i}^{\prime}\times{\bf S}_{i+\delta}^{\prime}\rangle\neq 0$.

An issue of coplanar magnets in an in-plane magnetic field is that if the magnetic order stays coplanar, or undergoes a certain flop transition that develops out-of-plane component. Numerous noncollinear magnets are known to undergo such flop transition, most notably in RMnO$_{3}$ (R$=$Gd, Tb, Eu, Y, or Ty)\cite{Kimura03,Kimura05,Yamasaki08,Murakawa08,Kida08,Kagawa09}, ZnCr$_{2}$Se$_{4},$\cite{Murakawa08_2} and DyFeO$_{3}$,\cite{Tokunaga08} where the flopping of magnetic order is indicated by measuring the electric polarization ${\bf P}_{ij}\propto{\hat{\bf e}}_{ij}\times\left({\bf S}_{i}\times{\bf S}_{j}\right)$\cite{Katsura05,Mostovoy06}. Here ${\hat{\bf e}}_{ij}$ is the vector that connects site $i$ and $j$. Below we show that the classical energy described by Eq.~(\ref{FM_spiral_energy_with_B_field}) already shows hints for a flop transition. Consider the spin ${\bf S}_{i}$ buckles out of plane for a small angle $\gamma_{i}$ from the coplanar state described by Eq.~(\ref{FM_spiral_energy_with_B_field}). The energy of this buckling state is 
\begin{eqnarray}
E^{\prime}&=&\sum_{i,\delta}{\tilde J}_{\delta}S^{2}\cos\gamma_{i}\cos\gamma_{i+\delta}\cos\left(\theta_{i+\delta}-\theta_{i}-\theta_{\delta}\right)
\nonumber \\
&+&JS^{2}\sin\gamma_{i}\sin\gamma_{i+\delta}-\sum_{i}g\mu_{B}BS\cos\gamma_{i}\cos\theta_{i}
\nonumber \\
&\approx& E-\sum_{i,\delta}{\tilde J}_{\delta}S^{2}\left(\frac{\gamma_{i}^{2}+\gamma_{i+\delta}^{2}}{2}\right)\cos\left(\theta_{i+\delta}-\theta_{i}-\theta_{\delta}\right)
\nonumber \\
&+&JS^{2}\gamma_{i}\gamma_{i+\delta}+\sum_{i}g\mu_{B}BS\frac{\gamma_{i}^{2}}{2}\cos\theta_{i}
\nonumber \\
&=&E+\Delta E
\end{eqnarray} 
where $E$ is that described in Eq.~(\ref{FM_spiral_energy_with_B_field}). It is clear that $\Delta E<0$ ($>0$) favors (disfavors) buckling. For long wave length spiral $D_{\delta}\ll|J|$, and writing $\theta_{i+\delta}-\theta_{i}=D_{f}/|J|\approx D_{\delta}/|J|\approx\theta_{\delta}$, the energy gain is 
\begin{eqnarray}
\Delta E&\approx&\sum_{i,\delta}|J|\frac{\left(\gamma_{i}-\gamma_{i+\delta}\right)^{2}}{2}
\nonumber \\
&+&\sum_{i,\delta}|J|\left(\frac{\gamma_{i}^{2}+\gamma_{i+\delta}^{2}}{2}\right)\frac{D_{f}}{|J|}\left(\frac{D_{\delta}}{|J|}-\frac{D_{f}}{2|J|}\right)
\nonumber \\
&+&\sum_{i}g\mu_{B}BS\frac{\gamma_{i}^{2}}{2}\cos\theta_{i}
\label{buckling_energy_gain}
\end{eqnarray}
The first two terms on the right hand side of Eq.~(\ref{buckling_energy_gain}) are positive, which disfavors bucking, but the third term due to magnetic field can be of either sign. Therefore in principle the spins can buckle out of plane to save energy at certain magnetic field strength, indicating a certain flop transition. Certainly to precisely model the flop transition for a particular compound, one needs to take into account the anisotropy due to crystalline symmetry. Nevertheless, our calculation indicates that the flop transition can be solely driven by the competition between DM interaction and the magnetic potential energy.

%We can solve the angle $\theta_{i}$ by minimizing Eq. (\ref{FM_spiral_energy_with_B_field}). A demonstrative example of $N=20$ sites 1D chain with DM interaction $D_{a}=0.4J$ is shown in Fig. \ref{fig:FM_spiral}, where one sees that at $g\mu_{B}|{\bf B}|\sim D_{a}$, a sizable spin current is already created. In this range of magnetic field, however, the spin current is not stable because the angular momentum transferred by the spin current causes the spins to be dynamic, and certain nonlinear behavior is expected. One has to go beyond Eq. (\ref{FM_spiral_energy_with_B_field}) and use two solid angles and the commutation relation in all three spin directions to capture the nonlinear dynamics of every spin in the chain, which is not the purpose of this study. 

The best way to produce a (nearly uniform and stable) spin current that can flow out of the sample is to apply a strong magnetic field $g\mu_{B}|{\bf B}|\gg D_{\delta}$, such that all the spins point along the direction of the magnetic field  $\theta_{i+\delta}=\theta_{i}=0$, as indicated in Fig.~\ref{fig:spiral_at_large_B}(b). The spin current on each bond is then 
\begin{eqnarray}
\lim_{g\mu_{B}B\gg D_{\delta}}J_{s}^{y}=|\tilde{J}|S^{2}\sin\theta_{\delta}=|\tilde{J}|S^{2}\frac{D_{\delta}}{|\tilde{J}|}=D_{\delta}S^{2}
\label{FM_spiral_in_B_current}
\end{eqnarray}
which only depends on the DM interaction and magnitude of the spins. Since the exchange interaction part of the classical energy(first term in Eq. (\ref{FM_spiral_energy_with_B_field})) is much smaller than the magnetic potential energy(second term in Eq. (\ref{FM_spiral_energy_with_B_field})) in this limit, one can disregard the small vibration or precession due to transfer of angular momentum, and the spin current is fairly stable. Note that buckling out of plane is disfavored in this limit, since $\theta_{i}=0$ makes the third term in Eq.~(\ref{buckling_energy_gain}) also positive, hence $\Delta E>0$. One reaches the conclusion that in both $g\mu_{B}|{\bf B}|=0$ and $g\mu_{B}|{\bf B}|\gg D_{\delta}$ limit the spins remain coplanar, although somewhere in between a flop transition occurs according to Eq.~(\ref{buckling_energy_gain}).

%{\cblue (1)Need to comment on the following: (a)When ${\bf B}$ field is on an arbitrary direction, not just ${\bf B}\perp{\bf D}$. One can still easily find the classical energy. (b)When the spiral order is not coplanar. Then what's the formula to calculate spin current? Is it still something like $J_{s}=-\partial E/\partial\theta$? Actually I'm not sure about this, maybe need to find some references. }

%\begin{figure}[ht]
%\begin{center}
%\includegraphics[clip=true,width=0.97\columnwidth]{FM_spiral.eps}
%\caption{ (color online) Bond-averaged spin current of a $N=20$ sites FM spiral chain in a magnetic field applied along the chain direction, with DM interaction $D_{a}=0.4J$. Inserts show  $\sin\left(\theta_{i+a}-\theta_{i}-\theta_{a}\right)\propto J_{s,i,i+a}$, which represents the spin current on each bond from $i=1\sim 19$, evaluated at $g\mu_{B}B/|J|=0.1$, $0.2$, and $0.8$. The blue arrows indicates configuration of first few spins near the open end. } 
%\label{fig:FM_spiral}
%\end{center}
%\end{figure}

The same analysis also applies to an AF spiral, as we demonstrate below. Consider Eq. (\ref{DM_Hamiltonian}) again but $J>0$, then $\tilde{J}_{\delta}=\sqrt{J^{2}+D_{\delta}^{2}}>0$, $\sin\theta_{\delta}=-D_{\delta}/\tilde{J}_{\delta}$, and Eq. (\ref{DM_Hamiltonian_2}) is still applicable. We parameterize the angle of the spins in the following way: we denote the magnitude of spin, as if it is a FM spiral, by $\langle {\bf S}^{FM}\rangle=S(\sin\theta_{i},0,\cos\theta_{i})$. The true AF spins are denoted by $\langle{\bf S}\rangle=S(\sin\tilde{\theta}_{i},0,\cos\tilde{\theta}_{i})$. Apparently $
\tilde{\theta}_{i}=\theta_{i}+{\rm Mod}\left[i,2\right]\cdot\pi$. In equilibrium, $\theta_{i+\delta}=\theta_{i}+\theta_{\delta}$, hence
\begin{eqnarray}
\tilde{\theta}_{i+\delta}&=&\tilde{\theta}_{i}+\left\{{\rm Mod}\left[i+\delta,2\right]-{\rm Mod}\left[i,2\right]\right\}\cdot\pi+\theta_{\delta}
\nonumber \\
&=&\tilde{\theta}_{i}+(-1)^{i}\pi+\theta_{\delta}
\end{eqnarray}
Together with an in-plane magnetic field ${\bf B}\perp{\bf D}$, the classical energy is
\begin{eqnarray}
E&=&\sum_{i,\delta}-\tilde{J}_{\delta}S^{2}\cos\left(\tilde{\theta}_{i+\delta}-\tilde{\theta}_{i}-(-1)^{i}\pi-\theta_{\delta}\right)
\nonumber \\
&-&\sum_{i}g\mu_{B}BS\cos\tilde{\theta}_{i}
\nonumber \\
&=&\sum_{i,\delta}-\tilde{J}_{\delta}S^{2}\cos\left(\theta_{i+\delta}-\theta_{i}-\theta_{\delta}\right)
\nonumber \\
&-&\sum_{i}g\mu_{B}BS(-1)^{i}\cos\theta_{i}
\label{E_AF_spiral}
\end{eqnarray}
On the basis of $\theta_{i}$, Eq. (\ref{E_AF_spiral}) has the same form as Eq. (\ref{FM_spiral_energy_with_B_field}) for a FM spiral, except the magnetic potential has an additional $(-1)^{i}$ factor due to alternating spins.

%Since an 1D AF chain is unstable, we start by considering a 2D square lattice Heisenberg model with Dzyaloshinskii-Moriya interaction, which contains a coplanar spiral order. A theory of this kind has been shown to be a candidate for underdoped cuprates{\cred(cite Sushkov)}, especially given the noncollinear order well explains the neutron scattering data{\cred(cite neutron data)}. Note that in the spiral theory of cuprates, the noncollinear order comes from a detail balance between kinetic energy of holes and the energy from spin texture. In contrast, our adoption for this model is purely for the sake of addressing field-induced spin current, so the Dzyaloshinskii-Moriya interaction can come from intrinsic parity breaking of the structure or some other mechanisms, and no charge degrees of freedom is involved in the following analysis. 

%{\cred (1)Need a formalism for AF spiral from Hamiltonian. (2)Need a 2D formalism as I argue the minimum model is a 2D AF spiral. }

%{\cblue (1)Maybe there's no need to do this calculation at all. Just say from the calculation of FM spiral, I know that for $B\leq J$ then I'll have spin current but the spin configuration is dynamical because of angular momentum conservation. So just say I need $B\ll J$ to get a static current. }

The spin current on a particular bond is 
\begin{eqnarray}
J_{s,i,i+\delta}^{y,AF}&=&-\frac{\partial E}{\partial (\tilde{\theta}_{i+\delta}-\tilde{\theta}_{i})}
\nonumber \\
&=&-\tilde{J}_{\delta}S^{2}\sin\left(\tilde{\theta}_{i+\delta}-\tilde{\theta}_{i}-(-1)^{i}\pi-\theta_{\delta}\right)\;.
\end{eqnarray}
Notice that we should use $\tilde{\theta}_{i}$ to calculate the spin current because it is the direction of the true spin. The polarization of the spin current again points out of plane. At small field, the angular momentum transfer due to spin current again makes the spin configuration to be dynamic. At $g\mu_{B}|{\bf B}|\gg J$, all spins align to $\tilde{\theta}_{i}\approx 0$, so 
\begin{eqnarray}
\lim_{g\mu_{B}B\gg J}J^{y,AF}_{s,\delta}=-\tilde{J}_{\delta}S^{2}\sin\left(-(-1)^{i}\pi-\theta_{\delta}\right)
=D_{\delta}S^{2}\;,
\end{eqnarray}
which is again only determined by DM interaction and magnitude of the spins, and is stable since all spins are pinned by the strong magnetic field. To reach this $g\mu_{B}|{\bf B}|\gg J$ limit is, however, practically impossible at present, since a typical value of exchange coupling $J\sim 0.1$eV implies a huge ${\bf B}$ field. Nevertheless, our calculation shows that this field-induced spin current is a robust feature regardless the sign of exchange coupling. The point is that as long as the coplanar magnet is lifted from its ground state by certain means (here by magnetic field) but the spins remain coplanar, the commutation relation $\left[\theta,S^{y}\right]=i$ ensures the existence of a spin supercurrent that can flow out of the system. For spin textures that are not confined in a plane, such as those created by a ${\bf B}$ field that has three-dimensional character\cite{Schutz03}, we anticipate that a spin supercurrent can also occur. In this case one has to use the full commutation relation $\left[S^{\alpha},S^{\beta}\right]=i\epsilon^{\alpha\beta\gamma}S^{\gamma}$ to calculate the spin supercurrent and work out the corresponding spin dynamics (if any), which requires further investigations.

A variety of multiferroics are known to display spiral, helical, or more complicated noncollinear order, including Tb$_{2}$Mn$_{2}$O$_{5}$\cite{Gardner88,Hur04}, Eu$_{1-x}$Y$_{x}$MnO$_{3}$\cite{Hemberger07}, Ni$_{3}$V$_{2}$O$_{8}$\cite{Lawes05},
LiCuVO$_{4}$\cite{Gibson04,Enderele05,Banks07,Schrettle08,Mourigal11}, LiCu$_{2}$O$_{2}$\cite{Matsuda04,Gippius04,Park07}, MnWO$_{4}$\cite{Taniguchi06}, TbMnO$_{3}$\cite{Kenzelmann05}, CoCr$_{2}$O$_{4}$\cite{Yamasaki06}, and BiFeO$_{3}$\cite{Andrzejewski13}. The interest on these materials partially comes from their connection to electric polarization ${\bf P}$, especially to test the interplay between magnetism and ferroelectricity. A trend of at least several of these materials is that in large field, typically $|{\bf B}|\sim 10$T, they indeed undergo a transition to (modulated) collinear order\cite{Schrettle08,Andrzejewski13}. Thus our rough estimation solely based on classical energy at least qualitatively captures the magnetic ordering at large field. The claim is that if the collinear state of these materials comes from applying a strong ${\bf B}$ field in a coplanar state, then it contains a uniform spin current up to the scale of correlation length.

The spin supercurrent can be varified by detecting the voltage drop associated with the induced electric dipolar field\cite{Meier03}. The typical spiral wave length is $\sim 100$ lattice constants, so $D\sim 1$meV if $J\sim 100$meV, which means $|{\bf B}|\sim 10$T can make all the spins parallel, consistent with that reported in Ref.~\onlinecite{Andrzejewski13}. According to Eq.~(\ref{FM_spiral_in_B_current}), for a wire of $\mu$m thickness and lattice constant $\sim$nm, the spin supercurrent produced in this parallel configuration is $\sim 10^{18}\mu_{B}/s$, hence a electric dipolar field\cite{Meier03} $|{\bf E}|\sim 10^{-2}$V/m at distance $r\sim 10^{-5}$m away from the surface of the wire. The voltage drop for two probes that are $\sim 10^{-5}$m apart is $10^{-7}$V, which is easily within the reach of current experimental technique.

%{\cred (1)Question: It seems like if the high-field collinear state has spin current, then a polarization ${\bf P}$ should occur according to Katsura-Nagaosa-Balatsky. But Shrettle08 says in this phase there is no ${\bf P}$. ----> But Peter said that's because  ${\bf P}\propto{\bf \hat{e}}_{ij}\times\left({\bf S}_{i}\times{\bf S}_{j}\right)$ according to Katsura-Nagaosa-Balatsky. Maybe it's consistent, because this material is described by $J_{1}-J_{2}$ model, so it seems right: the ground state is due to superexchange but frustrated so it shows spiral but ${\bf P}\neq 0$. Now I use magnetic field to kill spiral so no more ${\bf P}$ according to Katsura-Nagaosa-Balatsky. So it's ok. }

\section{conclusions}

In summary, we demonstrate that a dissipationless spin current flows between two coupled ferromagnets if their magnetizations are misaligned. On a phenomenological level, this principle is analogous to Josephson effect, in the sense that the dissipationless spin current is proportional to sine of the misalignment angle $J_{s}^{y}\propto\sin\theta$, which is a result of the commutation relation $\left[\theta,S^{y}\right]=i$ that is analogous to $\left[\phi,{\hat N}\right]=i$ in Josephson effect. The microscopic mechanisms for the dissipationless spin current depend on the nature of the two ferromagnets, whether they are metallic or insulating: In a FMM/I/FMM junction, it is due to proximity-induced phase gradient of the magnetization. In a FMM/FMI junction, the interface $s-d$ coupling between conduction electrons and localized spins yields a Landau-Lifshitz type of dynamics. In a FMI/SC/FMI or FMI/N/FMI junction, it is the combination of proximity-induced magnetization and the Landau-Lifshitz dynamics at the interfaces that generate the dissipationless spin current. On the other hand, it has been shown that in a bulk magnetic insulator, the spin supercurrent is related to the vector chirality that originates from virtual hopping of electrons. We emphasize that none of these mechanisms explicitly depends on scattering with impurities or phonons.

We further propose that the ${\bf B}$ field is an efficient handle to create the misaligned magnetizations. A static, uniform ${\bf B}$ field can be used to induce the magnetization in one of the two coupled systems, or it can be applied to a bulk magnetic insulator to affect the collinear or noncollinear magnetic order, which allows challenge in the detection of the dissipationless spin current. Several experimental techniques are proposed to indirectly measure the dissipationless spin current, including using polarization-sensitive probes to measure the proximity-induced magnetization in the FMM/I/FMM junction, measuring $T_{c}$ reduction in FMI/SC/FMI junction (already demonstrated in Ref.~\onlinecite{Deutscher69} and \onlinecite{Hauser69}), and comparing the electric polarization of multiferroics in the presence and in the absence of a ${\bf B}$ field. The misalignment principle and the use of a ${\bf B}$ field provide a practical way to generate dissipationless spin current on a mesoscopic scale, and indicate its application in device building. This possibility will be explored in a forthcoming study.

We thank H. Nakamura, S. Onoda, M. Mori, J. Mannhart, G. Logvenov, N. Nagaosa, A. Muramatsu, A. H. MacDonald, and S. Maekawa for stimulating discussions.

\end{document}